\documentclass[sigconf]{acmart}
\def\BibTeX{{\rm B\kern-.05em{\sc i\kern-.025em b}\kern-.08emT\kern-.1667em\lower.7ex\hbox{E}\kern-.125emX}}

\usepackage{color}
\usepackage{multirow}
\usepackage{subcaption}
\usepackage{graphicx}
\usepackage{hyperref}

\usepackage{eso-pic,xcolor}
\usepackage{caption}
\usepackage{graphicx}
\usepackage{verbatim}
\usepackage{subcaption}
\usepackage{multirow}
\usepackage{array}

\usepackage{enumitem}
\usepackage{siunitx}
\usepackage{amsmath}
\usepackage{amssymb}
\usepackage{mathtools}
\usepackage{fancyhdr}
    
\usepackage{graphicx}
\usepackage{color,soul}
\usepackage{xspace}

\renewcommand\footnotetextcopyrightpermission[1]{} % removes footnote with conference information in first column
\begin{document}

%
% The "title" command has an optional parameter, allowing the author to define a "short title" to be used in page headers.
\title{Experimenting with a Simulation Framework for Peer-to-Peer File Sharing in Named Data Networking}

%
% The "author" command and its associated commands are used to define the authors and their affiliations.
% Of note is the shared affiliation of the first two authors, and the "authornote" and "authornotemark" commands
% used to denote shared contribution to the research.

\author{Akshay Raman}
\affiliation{%
  \institution{University of California, Los Angeles}
  %\city{Los Angeles}
  %\country{USA}
  }
\email{akshay.raman@cs.ucla.edu}

\author{Kimberly Chou}
\affiliation{%
  \institution{University of California, Los Angeles}
  %\city{Los Angeles}
  %\country{USA}
  }
\email{klchou@cs.ucla.edu}

\author{Spyridon Mastorakis}
\affiliation{%
  \institution{University of Nebraska, Omaha}
  %\city{Omaha}
  %\country{USA}
  }
\email{smastorakis@unomaha.edu}

%\author{Lixia Zhang}
%\affiliation{%
%  \institution{UCLA}
%  \city{Los Angeles}
%\email{lixia@cs.ucla.edu}

%
% By default, the full list of authors will be used in the page headers. Often, this list is too long, and will overlap
% other information printed in the page headers. This command allows the author to define a more concise list
% of authors' names for this purpose.

%
% The abstract is a short summary of the work to be presented in the article.
\begin{abstract}
Peer-to-peer file sharing envisions a data-centric dissemination model, where files consisting of multiple data pieces can be shared from any peer that can offer the data or from multiple peers simultaneously. This aim, implemented at the application layer of the network architecture, matches with the objective of Named Data Networking (NDN), a proposed Internet architecture that features a data-centric communication model at the network layer. To study the impact of a data-centric network architecture on peer-to-peer file sharing, we proposed nTorrent, a peer-to-peer file sharing application on top of NDN. Since the initial nTorrent proposal in 2017, we have implemented its design in ndnSIM, the de facto NDN simulator. In this paper, we present the design of our nTorrent simulation framework, discussing various design decisions and trade-offs. We also describe our experimentation and validation process to ensure that our framework possesses the fundamental properties of nTorrent\let\thefootnote\relax\footnotetext{This paper is a preprint of~\cite{raman2019simulation}. The definitive version will be published by the ACM library.}. 
\end{abstract}

%\begin{CCSXML}
%<ccs2012>
%<concept>
%<concept_id>10003033.10003079.10003081</concept_id>
%<concept_desc>Networks~Network simulations</concept_desc>
%<concept_significance>500</concept_significance>
%</concept>
%<concept>
%<concept_id>10003033.10003079.10003082</concept_id>
%<concept_desc>Networks~Network experimentation</concept_desc>
%<concept_significance>300</concept_significance>
%</concept>
%<concept>
%<concept_id>10010147.10010341.10010366.10010369</concept_id>
%<concept_desc>Computing methodologies~Simulation %tools</concept_desc>
%<concept_significance>500</concept_significance>
%</concept>
%</ccs2012>
%\end{CCSXML}

%\ccsdesc[500]{Networks~Network simulations}
%\ccsdesc[300]{Networks~Network experimentation}
%\ccsdesc[500]{Computing methodologies~Simulation tools}

%
% Keywords. The author(s) should pick words that accurately describe the work being
% presented. Separate the keywords with commas.

%
% This command processes the author and affiliation and title information and builds
% the first part of the formatted document.
\maketitle

\section{Introduction}

Peer-to-peer file sharing applications, such as BitTorrent~\cite{BTrobust}, have been an area of active research over the years. In such applications, a file consists of multiple individual pieces of data. The primary goal of peers is to retrieve the desired data pieces from any other peer that can offer them. This data-centric functionality exists only at the application layer of the network architecture. Since peers operate on top of the point-to-point TCP/IP architecture, they need to decide from which peer (IP address) to retrieve the desired data. Specifically, peers have to discover other peers, select specific peers to establish TCP connections to, as well as estimate the quality of these connections.

Named Data Networking (NDN)~\cite{zhang2014named} is a proposed Internet architecture that offers a data-centric communication model. NDN enables applications to exclusively focus on what pieces of data to fetch, while the NDN network determines from where to fetch the desired data. In other words, NDN can provide the data-centric functionality needed by peer-to-peer applications directly at the network layer of the network architecture.

To explore the benefits and trade-offs of the NDN data-centric communication model for peer-to-peer file sharing, we have proposed nTorrent~\cite{mastorakis2017ntorrent}; an application for peer-to-peer file sharing in NDN. nTorrent was initially designed and evaluated in 2017. Since then, we have been working on implementing nTorrent in ndnSIM~\cite{mastorakis2017evolution}, the de facto NDN simulator, which has been designed and built on top of ns-3. Our goal has been to encourage and facilitate research and experimentation with ns-3, nTorrent and, NDN by providing an easy-to-use simulation framework for peer-to-peer file sharing (\emph{our framework is available at \url{https://github.com/AkshayRaman/scenario-ntorrent}}).

To help more people gain familiarity with our nTorrent simulation framework, we present our effort on implementing nTorrent in ndnSIM in this paper. We first describe our implementation design and its major components. We then present the validation process of our implementation through a set of simulation experiments. These experiments demonstrate that our implementation indeed possesses the desired features of the nTorrent design, including the ability to utilize multiple network paths for data downloading, to discover and retrieve data closer to the requesting peer, and to adapt to peer dynamics. We hope that the research community finds our framework useful, not only for experimentation with nTorrent, but also as the basis for the design, implementation, and evaluation of new peer-to-peer applications in ns-3 and NDN.

The rest of this paper is organized as follows: in Section~\ref{sec:back}, we present some brief background on NDN, the nTorrent design, and ndnSIM. We also discuss some prior related work. In Section~\ref{sec:des}, we present the design of our nTorrent simulation framework, while, in Section~\ref{sec:val}, we present our simulation validation. In Section~\ref{sec:lessons}, we discuss the lessons we learned while developing the simulation framework. Finally, Section~\ref{sec:concl} concludes our paper and discusses our future work.

\section{Background and Prior Work}
\label{sec:back}

In this section, we present a brief overview of the NDN architecture, the nTorrent application, and the ndnSIM simulation framework. We also discuss prior work on peer-to-peer network simulators. Our goal here is to help readers better comprehend what will be discussed in the rest of the paper.

\subsection{Named Data Networking}

The NDN~\cite{zhang2014named} architecture enables applications to retrieve data by its name based on a receiver-driven communication model. Consumer applications send requests, called \emph {Interest packets}, for the desired data, which are forwarded toward the producer(s) of the data based on their names by NDN Forwarding Daemons (NFDs)~\cite{nfd-dev}. Once an Interest reaches a node that has the requested data, \emph{a data packet} is sent back that follows the reverse network path of the corresponding Interest. 

\begin{figure}[h]
  \centering
  \includegraphics[scale=0.55]{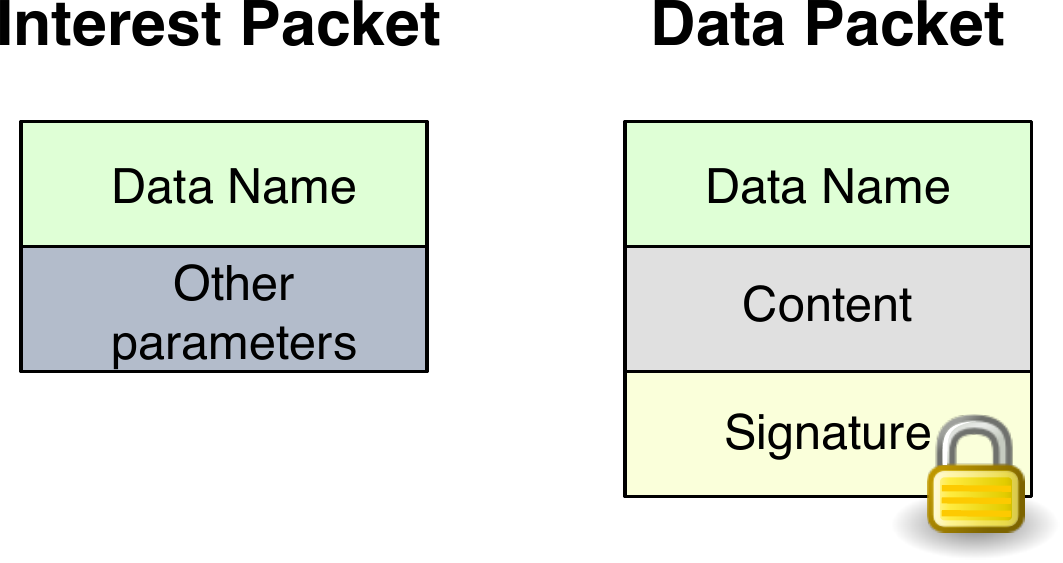}
  \caption{NDN Packet Format. An Interest Packet Consists of the Name of the Desired Data and other Parameters Used for its Forwarding by the NDN Network. A Data Packet Contains the Data Name, the Actual Content, and the Signature of the Data Producer}
  \label{Figure:ndn-packet-format}
  %\vspace{-0.5cm}
\end{figure}

As illustrated in Figure~\ref{Figure:ndn-packet-format}, an Interest packet consists of the name of the requested data and optional parameters. A data packet consists of the name of the data, the actual content, and a signature that binds the content to the name of the data packet~\cite{zhang2018overview, zhang2018security}. Each data packet is named and signed by its producer.

NFD forwards Interests and data based on three data structures. The first one, called Forwarding Information Base (FIB), contains a number of name prefixes along with the outgoing interfaces for each one and is used to forward Interests based on their name. The second one, called Pending Interest Table (PIT), contains network state for the Interests that have been forwarded, but the corresponding data has yet to be received. The third one, called Content Store (CS), acts as an in-network cache that stores recently retrieved data packets to satisfy future Interests for the same data, without the need for these Interests to be forwarded all the way to the producer.

Each NFD also includes a \emph{forwarding strategy} module, which decides whether, when, and through which outgoing face(s) to forward received Interests~\cite{chan2017fuzzy, mastorakis2018experimentation}. For example, after a FIB lookup is performed, it may be determined that an Interest can be forwarded through more than one interfaces, meaning that multiple network paths to the data may exist. In such a case, the forwarding strategy can forward an Interest: (i) through multiple interfaces at the same time to find the network path with the lowest latency to the data, or (ii) through one interface first and through alternative interfaces later, if the Interest fails to bring data back through the first interface (e.g., due to a link or a node failure). The forwarding strategy along with the network state maintained by each NFD through PIT enable the NDN forwarding plane to adapt to the network conditions and perform data retrieval through multiple different network paths.

\subsection{nTorrent: Peer-to-Peer File Sharing in NDN}
\label{ntorrent}

nTorrent~\cite{mastorakis2017ntorrent, mastorakis2019peer, mastorakis2017bit} is an application that provides peer-to-peer file sharing functions in NDN. Unlike BitTorrent that is implemented as an application layer overlay on top of the point-to-point TCP/IP architecture, nTorrent takes advantage of NDN that provides data distribution functions directly at the network layer of the network architecture. Furthermore, each network layer packet in NDN comes with its own signature, offering fine-grained verification of the retrieved data by nTorrent peers compared to the coarse-grained BitTorrent cryptographic hashes per piece (each piece typically consists of multiple network layer packets).

In nTorrent, each peer acts as a data producer and a data consumer at the same time; it downloads data from other peers or from in-network caches (data consumer) and shares its data with other peers (data producer). nTorrent takes advantage of NDN's forwarding strategy module to discover and retrieve data from the location (peers or in-network caches) closest to the requester that can provide the data. This allows nTorrent to minimize the generated network traffic, as well as adapt to peer dynamics, since new peers might join and existing peers might leave the swarm without a notice. 

\begin{figure}[h]
  \centering
  \includegraphics[width=\linewidth]{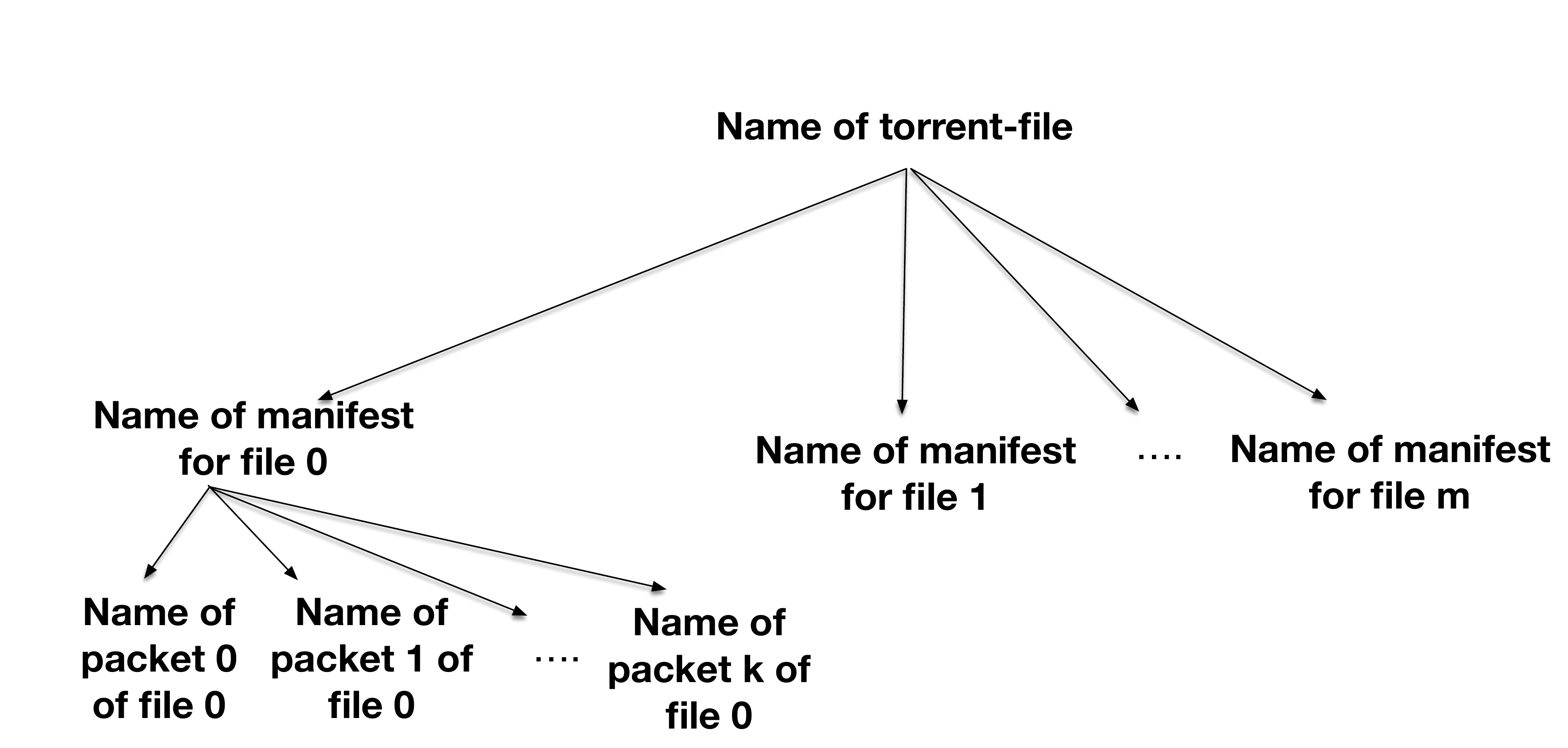}
  \caption{Hierarchical Process for Peers to Learn the Names of the Data Packets to Request for a Torrent. Peers First Retrieve the Torrent-File and then the Manifest of each File in the Torrent. A Manifest Contains the Names of the Individual Data Packets of a File in the Torrent}
  \label{Figure:hier-man}
\end{figure}

Each torrent in nTorrent may consist of multiple individual files and each file of multiple network layer NDN data packets. To learn the names of the data packets to request for the retrieval of a torrent, peers leverage a hierarchical process (Figure~\ref{Figure:hier-man}). Similar to BitTorrent, they first download a \emph{torrent-file} (peers learns the name of the torrent-file through an out-of-band mechanism, such as a website). In nTorrent, a torrent-file contains the names of one or more \emph{file manifests} (or manifests for short). A manifest contains the names of the data packets to request to download a file in the torrent. In other words, for each individual file in the torrent, there is a manifest that contains the names of the data packets in this file. As a result, after downloading a torrent-file and learning the names of the manifests, a peer can download each manifest to learn the names of the packets in each file of the torrent. Note that a torrent-file and the corresponding manifests are generated and signed by the peer that generated the torrent.

\subsection {ndnSIM: an Open-Source NDN Simulation Framework Based on ns-3}

ndnSIM~\cite{mastorakis2017evolution, mastorakis2015ndnsim, mastorakis2016ndnsim} is the de facto NDN simulator, which is based on ns-3. The first version of the simulator was released in 2012. Since then, ndnSIM has attracted hundreds of users from institutions all around the world. ndnSIM features a number of plug-and-play simulation scenarios, a website (\url{https://ndnsim.net/current/}) with detailed instructions on how users can experiment with the simulator, and an active mailing list (\url{https://www.lists.cs.ucla.edu/mailman/listinfo/ndnsim}) with over 550 subscribers.

ndnSIM is implemented as a (not yet merged) module of ns-3, making use of all the abstractions and the simulation environment offered by ns-3 (e.g., Node, Channel, Application, NetDevice abstractions). The latest versions of ndnSIM offer integration with the real-world NDN prototypes: NFD and the ndn-cxx library, which provides the abstractions for the basic NDN operations (e.g., packet encoding/decoding, security and packet transmission operations).

\begin{figure}[h]
  \centering
  \includegraphics[width=\linewidth]{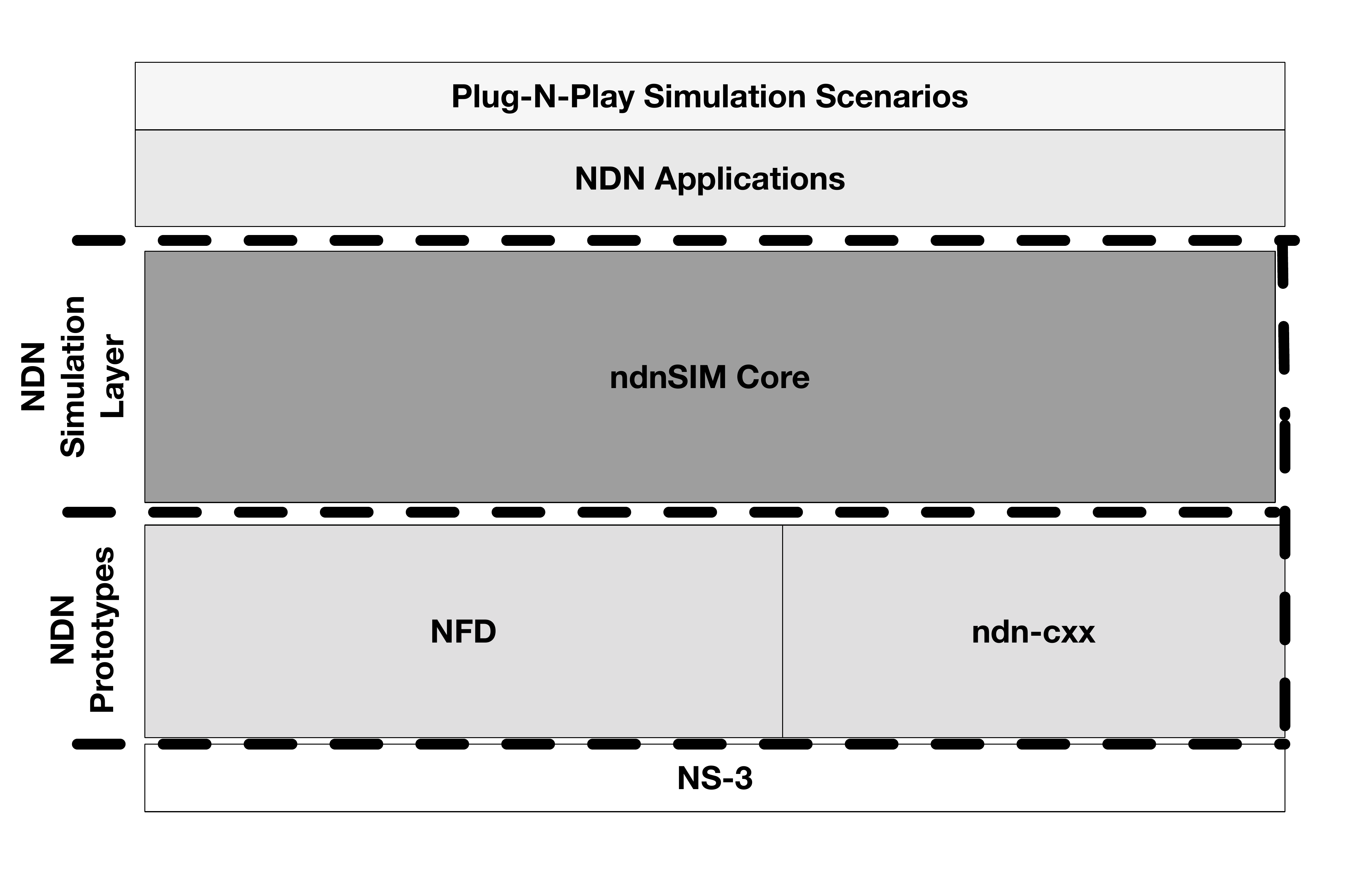}
  %\vspace{-1cm}
  \caption{The Structure of the ndnSIM Framework. ndnSIM is based on ns-3 and Offers Integration with NFD and ndn-cxx, as well as a Set of Pre-installed NDN Applications and Simulation Scenarios}
  \label{Figure:ndnsim}
\end{figure}

The overall structure of ndnSIM is illustrated in Figure~\ref{Figure:ndnsim}. The base of the entire framework is ns-3, offering the core simulation environment. On top of ns-3, there are the NDN prototypes, which are integrated with the NDN simulation layer (ndnSIM core) to power the NDN simulations for high fidelity of simulation results. Users can take advantage of pre-installed NDN applications or develop their own applications based on ns-3's Application abstraction. %Users can also develop or port (with minor modifications) to ndnSIM real-world NDN applications written based on the ndn-cxx library. This offers a ``two-way'' of experimentation and evaluation; real-world applications can be easily ported and simulated in ndnSIM and simulated applications based on the ndn-cxx library can be easily used in real-world deployments. 
Finally, users can run one of the tutorial simulation scenarios that come with ndnSIM or develop their own scenarios in the same way as they would create a simulation scenario in ns-3.

Some examples of use-cases that ndnSIM has been used for their simulation include the following: (i) schemes for distributed~\cite{krol2019compute} and edge computing~\cite{mastorakis2019towards, leecase}, (ii) distributed synchronization protocols~\cite{de2017design, li2018data, lidistributed}, (iii) mobility~\cite{zhang2018kite}, and (iv) link layer reliability~\cite{vusirikala2016hop}. 

\subsection{Prior Work on Peer-to-Peer Network Simulators}
\label{sec:related}

The most popular protocol for peer-to-peer file sharing in TCP/IP, BitTorrent~\cite{BTrobust}, has been improved and tested on multiple simulators. The simulation framework most relevant to our work is VODSim~\cite{weingartner2012building}, an implementation of BitTorrent for ns-3. In terms of design, VODSim is limited to the application-layer of TCP/IP, while our framework extends the stateful NDN forwarding plane to deal with peer dynamics (Section~\ref{strategy}). VODSim provides scripting abstractions for the easy setup of simulation scenarios, a feature that is not currently supported by our framework. We plan to add such a feature as a part of our future work (Section~\ref{sec:concl}).

LEDBAT~\cite{LEDBAT}, a congestion control protocol for BitTorrent, was extensively tested in ns-2, while Barcellos et al. deployed TorrentLab~\cite{TorrentLab}, a testbed to run BitTorrent related simulations and live experiments. Further research has been conducted to improve the BitTorrent robustness and performance~\cite{BTperf2004, BTperf2005, BTimprovesimple} through the use of simulations.

Research has been conducted to create new simulators for peer-to-peer data sharing protocols including 3LS~\cite{3LSsim}, PDST~\cite{PDSTsim}, and PeerfactSim.KOM~\cite{PeerfactSimKOM}. 3LS is used to study complex peer-to-peer networks. PDST focused on simulating peer-to-peer networks and peer database systems. PeerfactSim.KOM is a simulation framework that focused on gathering results about the performance of Gnutella.

Researchers have also studied peer-to-peer network scenarios in NDN. Specifically, they have studied the effect of user-assisted in-network caching on the performance of the file sharing process~\cite{ICNCache2013, ICNUserAssistedCaching, mastorakis2018real}. They have also proposed a routing protocol to manage the delivery of data~\cite{PeerAssistedRouting} and a system to synchronize files among multiple parties~\cite{FileSyncNDN}. 

As previous work has indicated, peer-to-peer networking has been an active research area in TCP/IP-based networks over the previous years. Through the implementation of a simulation framework for nTorrent, we aimed to provide to the broader research community a useful tool for experimentation with peer-to-peer applications in NDN.

% Google Doc Link of paper titles
% https://docs.google.com/document/d/1OtaqmmkBzV_IGXuVIAObtvj3sGhLyI2mzlJoeinfQXg/edit?usp=sharing

\section{nTorrent Simulation Framework Design}
\label{sec:des}

On a high level, the design of the nTorrent framework consists of three major components: (i) an nTorrent base library that contains commonly used abstractions (e.g., a class implementation of the torrent-file and related methods), (ii) two types of simulated applications, namely an nTorrent producer and an nTorrent consumer, and (iii) a forwarding strategy (network-layer component) that provides support to peer dynamics and enables data fetching from the closest (in terms of network delay) peer to the requester, so that the data retrieval latency is minimized. These components can be used and/or extended for the development of new peer-to-peer applications in the future.

The nTorrent producer application is analogous to a BitTorrent seeder, while nTorrent consumer application is analogous to a BitTorrent leecher. Specifically, the producer generates the torrent data, the torrent-file, and the manifests. The consumers have the goal of downloading the torrent and share the torrent data with other consumers as they download it.

Note that our framework can be combined with existing ns-3 models to create more complex experiments. For example, it can be combined with the Wi-Fi and LTE models to setup wireless file sharing experiments, as well as the energy and MPI models for energy measurements and parallelization of the simulation execution respectively. These models can be instantiated along with the nTorrent model in an ns-3 simulation scenario.

In the rest of this section, we present a communication use-case between two simulated nodes to explain the nTorrent network model. We then discuss each one of the major components of our framework.

\subsection {nTorrent Network Model}

\begin{figure}[h]
  \centering
  \includegraphics[scale=0.24]{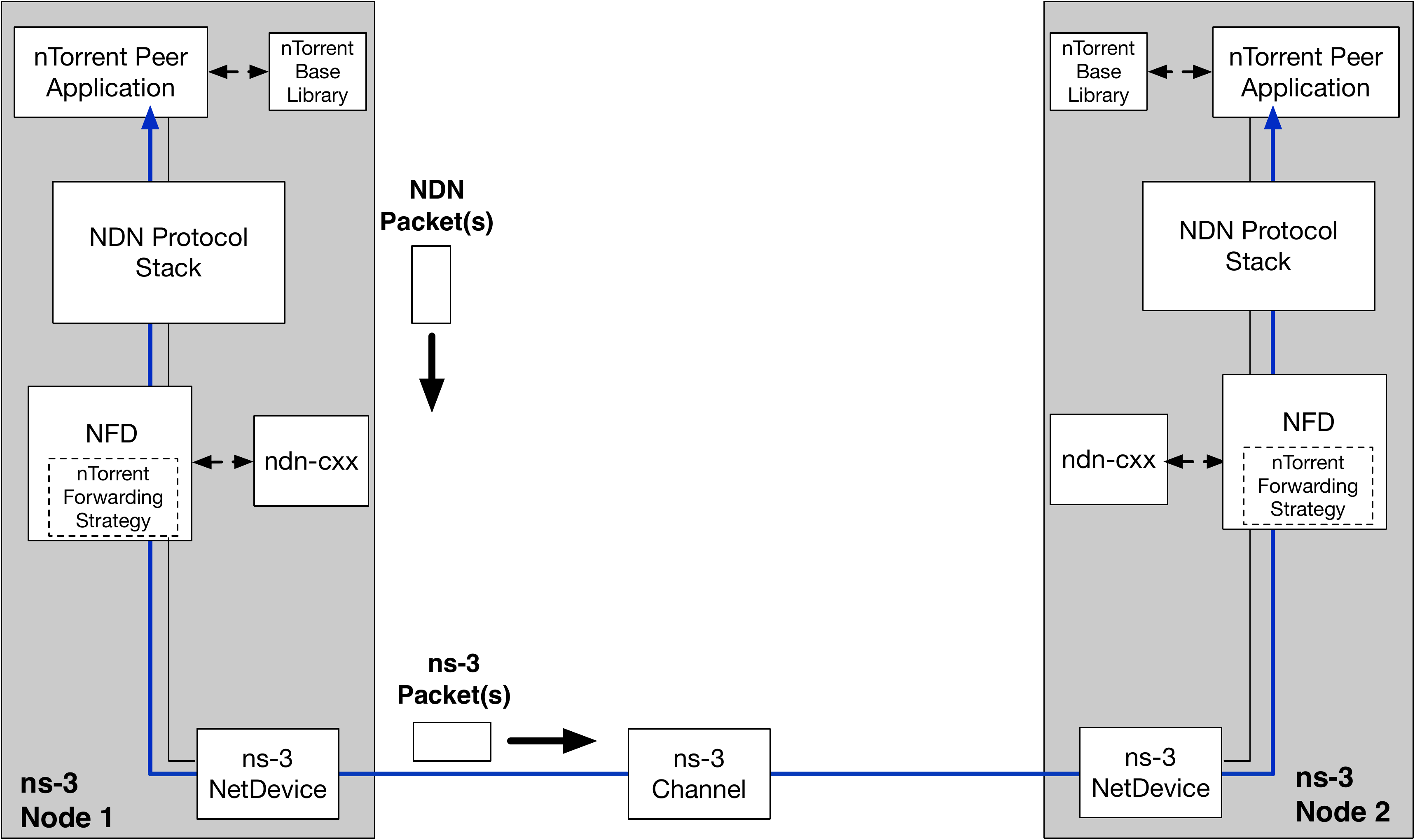}
  \centering
  \caption{nTorrent Network Model}
  \label{Figure:model}
  %\vspace{-0.5cm}
\end{figure}

In Figure~\ref{Figure:model}, we present a communication use-case between two simulated nodes running nTorrent. We assume that the NFD instance on each node runs the nTorrent forwarding strategy, which makes forwarding decisions about the nTorrent traffic. An nTorrent application instance running on Node 1 makes use of the nTorrent base library abstractions to generates NDN packets. These packets pass through NFD and the nTorrent forwarding strategy to an ns-3 NetDevice running on Node 1. Each NDN packet is converted to an ns-3 packet and is forwarded through an ns-3 Channel instance to Node 2. At Node 2, the ns-3 packet is converted back to an NDN packet and is forwarded by NFD to an nTorrent application running on this node.

This model relies on both NDN and ns-3 related abstractions and variables (e.g., the size of NDN caches, the size of ns-3 queues). It also introduces a number of new variables related to: (i) higher-layer semantics, such as segmentation of a torrent into individual NDN data packets (e.g., how many data packets, what is the size of each packet), (ii) the size of the generated torrent-file and manifests, (iii) Interest sending rates by peers for torrent retrieval, and (iv) forwarding strategy semantics (e.g., Interest satisfaction rate, other statistics as mentioned in Section~\ref{strategy}).

\begin{figure*}[h]
  \centering
  \includegraphics[scale=0.3]{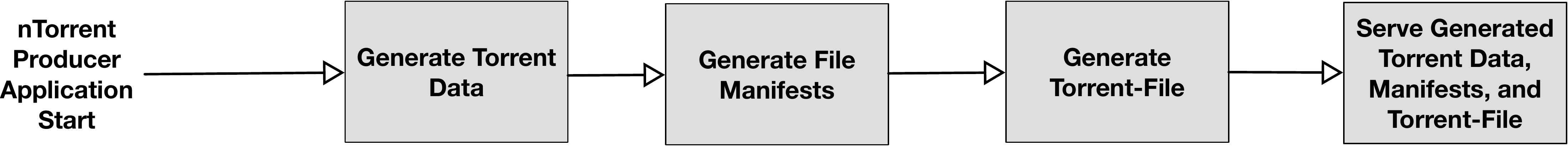}
  \centering
  \caption{Workflow of an nTorrent Producer Application}
  \label{Figure:producer}
  %\vspace{-0.5cm}
\end{figure*}

\subsection{nTorrent Base Library}
\label{library}

The base library contains a number of common abstractions and functions used by the consumer and producer applications. 

The library contains a \textcolor{blue}{TorrentFile} class. This class contains methods to generate a torrent-file, encode/decode a torrent-file object to/from the NDN packet format, add the name of a file manifest to the torrent-file, and remove the name of a manifest from it.

The \textcolor{blue}{FileManifest} class is a realization of the file manifest structure. Similar to the methods of the \textcolor{blue}{TorrentFile} class, it includes methods to generate a file manifest, encode/decode a file manifest to/from the NDN packet format, and add/remove names of data packets to/from a manifest.

To keep track of what torrent data a peer has received so far and what data it is still missing, we have implemented the \textcolor{blue}{TorrentManager} class. This class implements methods to identify by name if a peer has downloaded and can serve a requested data packet, as well as the name of the data packets that a peer is still missing and should download next.  

The library also includes: (i) a \textcolor{blue}{StatsTable} class, which is used by peers to collect statistics about the data downloading process (e.g., average latency, estimated download completion time) and (ii) an \textcolor{blue}{InterestQueue} class, which implements an application-layer queue for pacing the Interests sent by each peer. These abstractions are not currently used by the nTorrent producer and consumer applications. We plan to integrate them in the future to enhance the application flexibility. We also feel that these abstractions will be useful to new peer-to-peer applications that will be developed in the future.

\subsection{nTorrent Producer Application}

The workflow of the producer operation is illustrated in Figure~\ref{Figure:producer}. The \textcolor{blue}{NTorrentProducer} class uses methods of the \textcolor{blue}{TorrentManager}, \textcolor{blue}{TorrentFile}, and \textcolor{blue}{FileManifest} classes from the base library to generate the data packets for the torrent, the torrent-file, and the file manifests. 

For simulation purposes, the torrent data packets contain arbitrary data, rather than the actual data of real files (parsing and packetization of real files is also supported by the base library). The producer application accepts input from the simulation scenario that specifies the name of the torrent, the number of files in the torrent, and the number and size of packets for each file in the torrent. In this way, the application determines the total number of the torrent data packets to be generated. Subsequently, the producer application generates a file manifest for each file in the torrent, and, eventually, the torrent-file.

Once the producer has generated the required data, it starts serving requests sent from other peers. When the producer receives an Interest, it first checks the type of the Interest received (i.e., Interest for a torrent-file, a manifest, or a torrent data packet), and then responds with the corresponding data. 

%Some of the attributes of the \textcolor{blue}{NTorrentProducer} class include \textcolor{blue}{namesPerSegment}, \textcolor{blue}{namesPerManifest} and \textcolor{blue}{dataPacketSize}. \textcolor{blue}{namesPerSegment} refers to the number of names (files) per segment. \textcolor{blue}{namesPerManifest} refers to the number of names per manifest. The \textcolor{blue}{dataPacketSize} refers to the size of each data packet that is being simulated. All of these parameters can be modified at run-time using command line arguments. 

\subsection{nTorrent Consumer Application}
\label{sec:consumer}

\begin{figure}[h]
  \centering
  \includegraphics[scale=0.24]{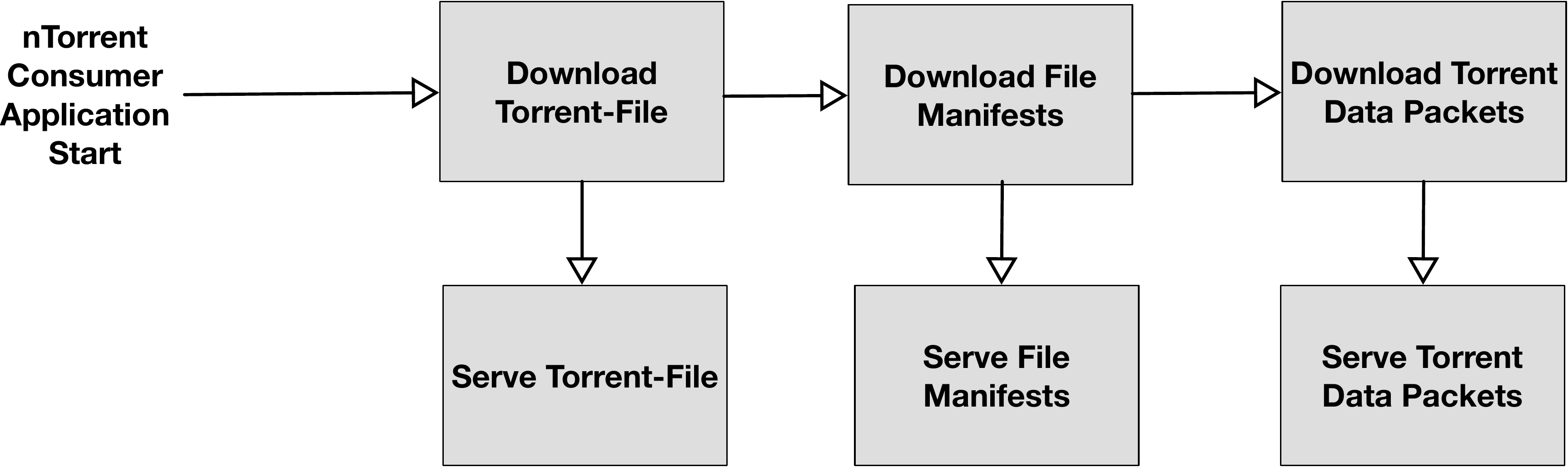}
  \caption{Workflow of an nTorrent Consumer Application}
  \label{Figure:consumer}
  %\vspace{-0.5cm}
\end{figure}

The workflow of the consumer operation is illustrated in Figure~\ref{Figure:consumer}. The \textcolor{blue}{NTorrentConsumer} class implements methods, so that a peer first sends Interests for the torrent-file. Once the torrent-file is retrieved, its content will be decoded to learn the names of the file manifests to request. Once the file manifests are requested and retrieved, a peer learns the names of each individual data packet of each file in the torrent. The peer requests these packets sequentially, starting from the first packet of the first file in the torrent to the last packet of the last file in the torrent. This data retrieval strategy has been proved to increase the utilization of the NDN in-network caches when multiple peers download the same torrent at the same time~\cite{mastorakis2017ntorrent, mastorakis2019peer}.

A consumer peer stores all the data it receives, including the torrent-file and file manifests, with the goal of serving it to others. To achieve that, a peer needs to inform the NDN network about the data names that it can serve. This is done with the interaction of the consumer application with a name-based routing protocol. Specifically, a peer announces the name prefix of the data it has to the routing protocol and the routing protocol propagates this announcement across the NDN network. As a result, a FIB entry is created for the announced name prefix at every NFD in the network, so that Interests from other peers can reach the peer that has the data. If a FIB entry for an announced name prefix already exists at an NFD, a new outgoing interface is added to this entry to denote the existence of a new network path for data retrieval.

\subsection{Forwarding Strategy}
\label{strategy}

Given the peer-to-peer nature of nTorrent, new peers might join or existing peers might leave without notice. To this end, we have designed and implemented a forwarding strategy, which is installed at every simulated NDN router in the network. This strategy takes advantage of the stateful NDN forwarding plane to deal with peer dynamics. Our strategy also aims to minimize the latency for the retrieval of the data by discovering the closest peer to the requester (in terms of network delay).

Our strategy, running at every NFD instance, collects statistics about the number of Interests that were able to bring a data packet back through each outgoing interface (Interest satisfaction rate), as well the average delay for data retrieval through each interface. The strategy instance takes advantage of the collected statistics to identify: 

\begin {itemize}[leftmargin=*]

\item When the selected outgoing interface cannot bring data back anymore (e.g., because a peer left the swarm). This is achieved by monitoring the Interest satisfaction rate $s$. When $s$ falls below a certain threshold ($s$ \textless $threshold$) or a number of consecutive Interests cannot retrieve data, the strategy selects an alternative outgoing interface to forward future Interests.

\item When an alternative interface can result in lower data retrieval latency $d$ than the one currently used for Interest forwarding, called $current\_interface$. This is achieved by probing alternative interfaces (e.g., by forwarding 1 out of every 50 Interests through an interface, called $probed\_interface$, other than $current\_interface$). If the latency through the probed interface $d_{probed\_interface}$ is lower than the latency through the current face $d_{current\_interface}$ ($d_{probed\_interface}$ \textless $d_{current\_interface}$), the strategy selects \\ $probed\_interface$ for the forwarding of future Interests.

\end {itemize}

This adaptive behavior of our strategy is based on the assumption that multiple outgoing interfaces exist at an NFD for a given name. If this is not case or an NFD has tried all the available interfaces without being able to retrieve data, it can send a Negative ACKnowledgement (NACK) to the previous hop (downstream) NFD. A NACK indicates the name of the Interest that was not able to be forwarded. Once the downstream NFD receives a NACK, our strategy module will be called to forward the Interest, whose name is specified in the received NACK, through another interface. 

Note that our strategy aims to utilize all the available links starting from the one that offers the lowest latency to the requested data, which we call \emph {primary link}. Once the primary link is fully utilized (i.e., its utilization has reached its bandwidth capacity\footnote{The forwarding strategy can identify when the utilization of a link starts reaching its full capacity by observing longer data retrieval delays than usual or even losses, since the queue of a node (droptail queue model of ns-3) gets longer, overflows, and eventually drops packets.}), our strategy expands Interest forwarding to the link that offers the second lowest latency to the data, while keeping the primary link fully utilized, and so on and so forth. As a result, the strategy is able to utilize all the available paths over the network for data retrieval, starting with the one that offers the lowest latency to the data, as we will demonstrate in Section~\ref{sec:results}.

\section{nTorrent Simulation Framework Validation}
\label{sec:val}

In this section, our goal is to validate through simulation experiments that our implementation can indeed provide the features of the nTorrent design. These features include:

\begin {itemize} [leftmargin=*]

\item Utilization of multiple network paths (if available) and maximization of download speed.

\item Discovery and retrieval of data closer (in terms of network latency) to the requester.

\item Adaptability to peer dynamics; peers can leave the swarm without notice, therefore, data needs to be retrieved from peers that are still available.  

\end {itemize}

% The objective of this paper is to implement a simulation framework for the nTorrent library as described in the nTorrent paper \cite{mastorakis2017ntorrent}. In this section, we talk about the metrics we used to validate the simulation. We aim to showcase the benefits of the inherent peer-to-peer nature of NDN using nTorrent. We have defined some sample topologies to highlight this feature of nTorrent. Some of the properties of a torrent setup include peer discovery, peer selection, data sharing incentivization, piece integrity verification and traffic localization \cite{mastorakis2017ntorrent}.

\subsection {Simulation Setup}
\label{sec:setup}

To validate our implementation, we implemented three simulation scenarios for the distribution of a torrent of size 100MB (each torrent data packet carries 1KB of torrent data). We run each scenario 10 times and we present the 90th percentile of the collected results.

\begin{figure}[h]
  \centering
  \includegraphics[scale=0.2]{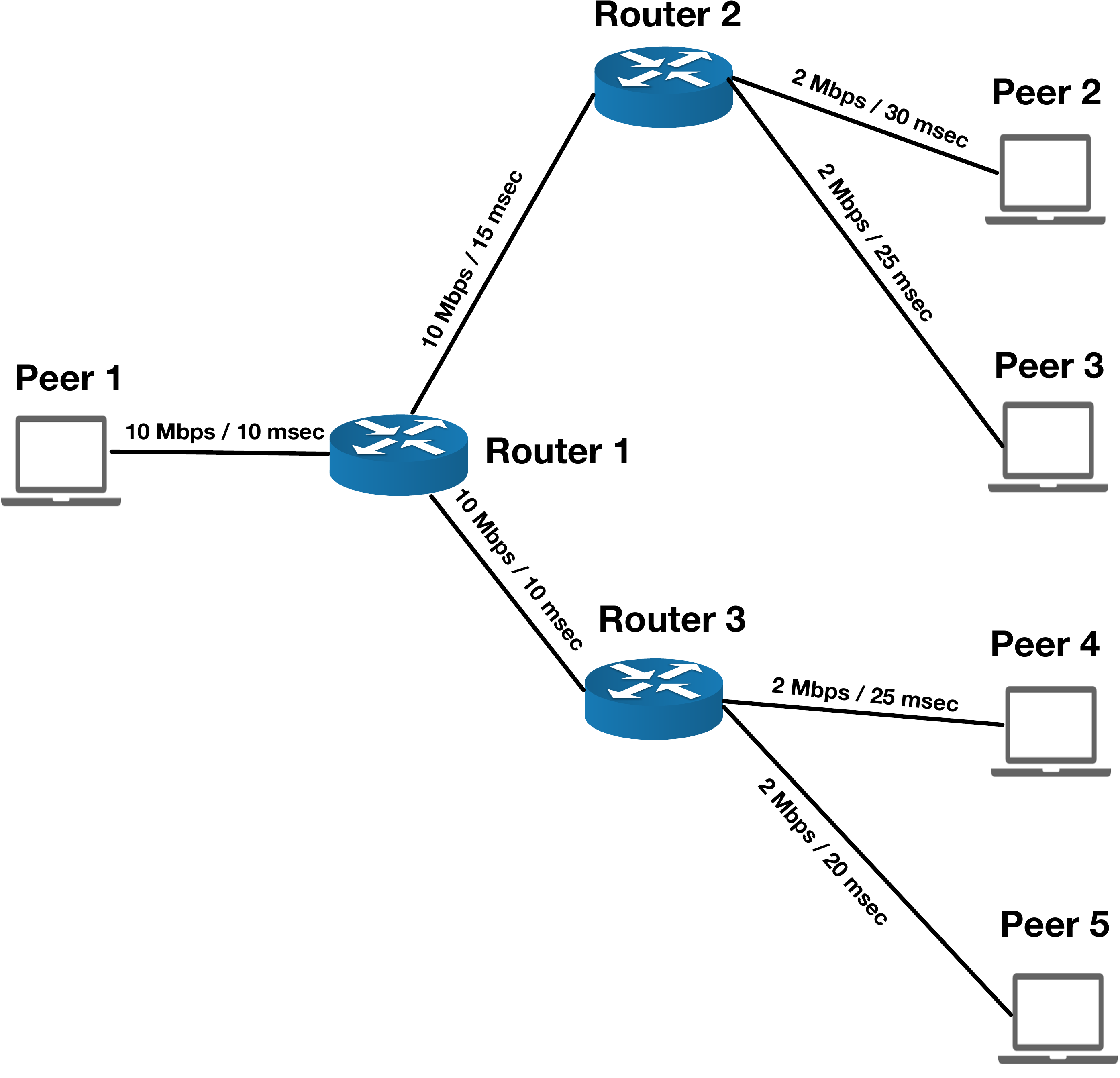}
  \centering
  \caption{Topology Used to Validate that nTorrent can Utilize Multiple Network Paths to Peers that can Provide the Torrent. The Topology Includes Fast (10Mbps) and Slow (2Mbps) Links and each Network Path has a Different Delay}
  \label{Figure:topology_routernodedeg3}
  %\vspace{-0.5cm}
\end{figure}

\textbf {Scenario 1:} We used the network topology of Figure~\ref{Figure:topology_routernodedeg3}, which includes links of different bandwidth, to validate that our implementation can sufficiently utilize multiple network paths to retrieve torrent data from multiple peers in parallel and maximize the speed of the downloading process. Specifically, peers 2, 3, 4, and 5 act as producers (seeders), having all the data for a torrent, while peer 1 acts as a consumer (leecher) that tries to download this torrent.

\begin{figure}[h]
  \centering
  \includegraphics[scale=0.2]{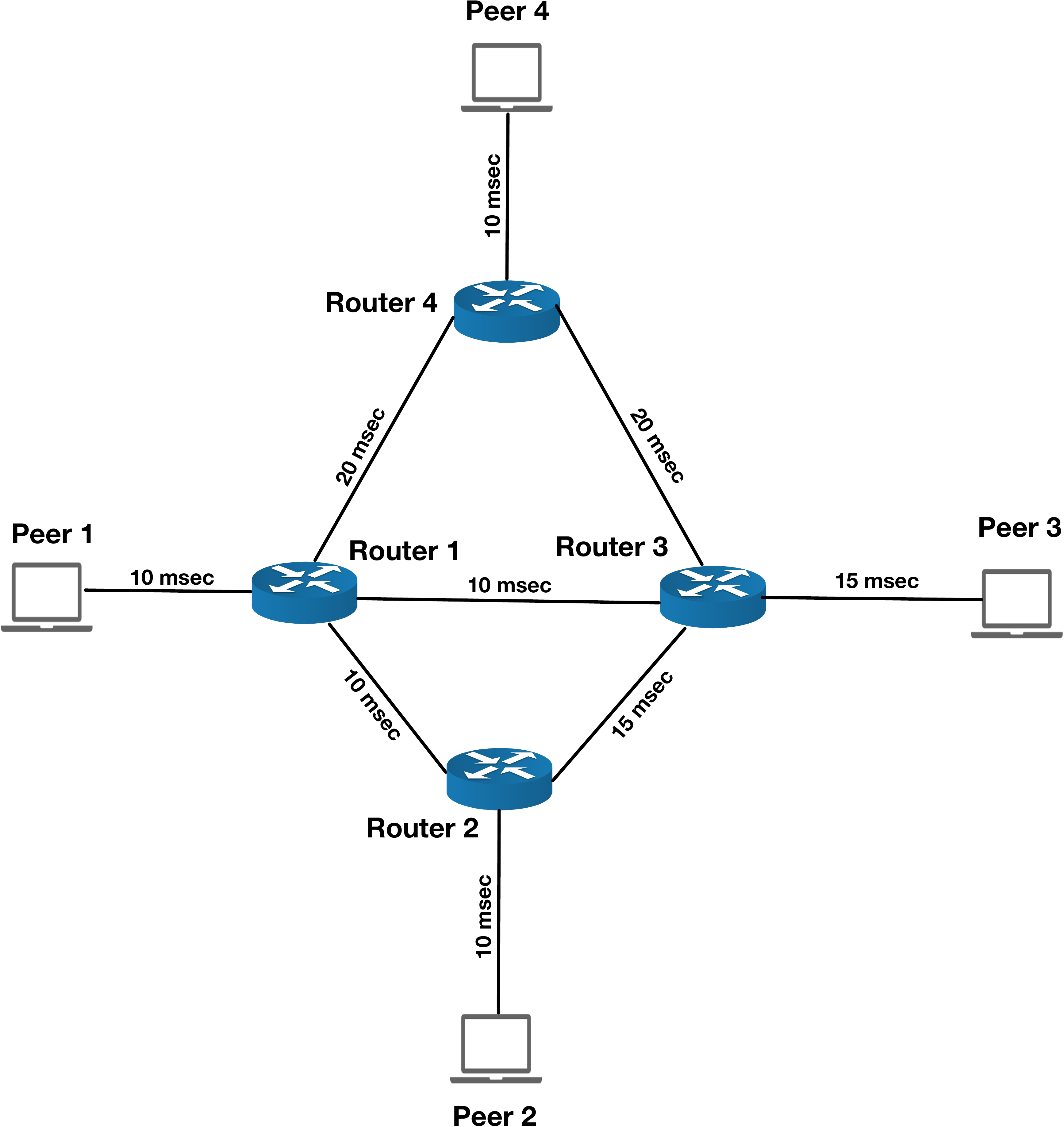}
  \centering
  \caption{Topology Used to Validate that nTorrent: (i) Can Discover Data with a Shorter Retrieval Latency to the Requester as the Data Becomes Available and, (ii) Can Adapt to Peer Dynamics. This Topology Includes Links with Different Delays without Bandwidth Bottlenecks}
  \label{Figure:topology_routernodedeg4}
  %\vspace{-0.5cm}
\end{figure}

\textbf {Scenario 2:} We used the network topology of Figure~\ref{Figure:topology_routernodedeg4} (no bottleneck links) to verify the discovery and retrieval of data closer to the requester. peer 4 has the torrent, acting as the torrent producer, and peers 1, 2, and 3 try to download the torrent. Peer 1 starts first at simulation time $t_1=0 sec$, while peer 2 starts at time $t_2=20 sec$, and peer 3 starts at time $t_3=40 sec$.

\textbf {Scenario 3:} We used the network topology of Figure~\ref{Figure:topology_routernodedeg4} (no bottleneck links) to verify the adaptation of nTorrent to peer dynamics. Specifically, peers 2, 3, and 4 have the torrent, while peer 1 starts downloading the torrent at simulation time $t_0=0 sec$. At time $t_1=20 sec$, peer 2 disconnects and peer 3 disconnects at time $t_2=40 sec$.

The routers of both Figures~\ref{Figure:topology_routernodedeg3} and~\ref{Figure:topology_routernodedeg4} run the nTorrent forwarding strategy presented in Section~\ref{strategy}. A name-based routing protocol is used to propagate the announcements of peers for torrent data name prefixes and create FIB entries across the network.

Note that we focus on the behavior of nTorrent, without trying to optimize its performance. We rather aim to fully understand whether our implementation can achieve the desired properties. To this end, we have disabled the NDN in-network caching feature, so that NDN routers do not cache torrent data. It has been demonstrated in~\cite{mastorakis2017ntorrent} that in-network caching can boost the nTorrent performance when multiple peers request the same torrent at the same time.

\subsection {Simulation Results}
\label{sec:results}

\begin{figure*}
    %\vspace{-0.5cm}
	\centering
	\begin{subfigure}{.33\textwidth}
		\centering
		\includegraphics[scale=0.20]{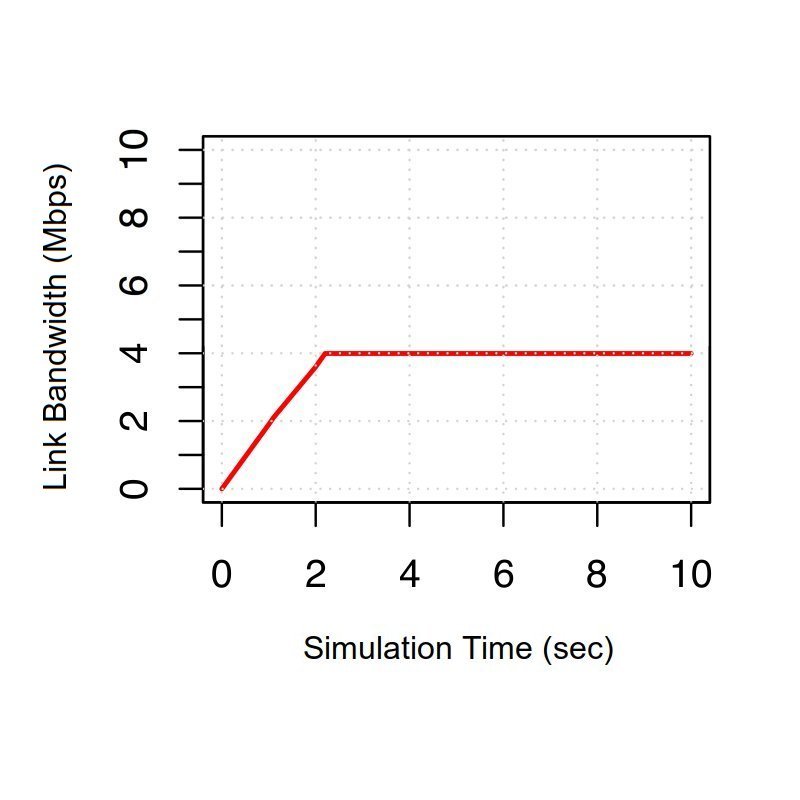}
		%\vspace{-0.8cm}
		\caption{Link between router 1 and router 3}
		\label{Figure:band1}
	\end{subfigure}
	\begin{subfigure}{.33\textwidth}
		\centering
		\includegraphics[scale=0.20]{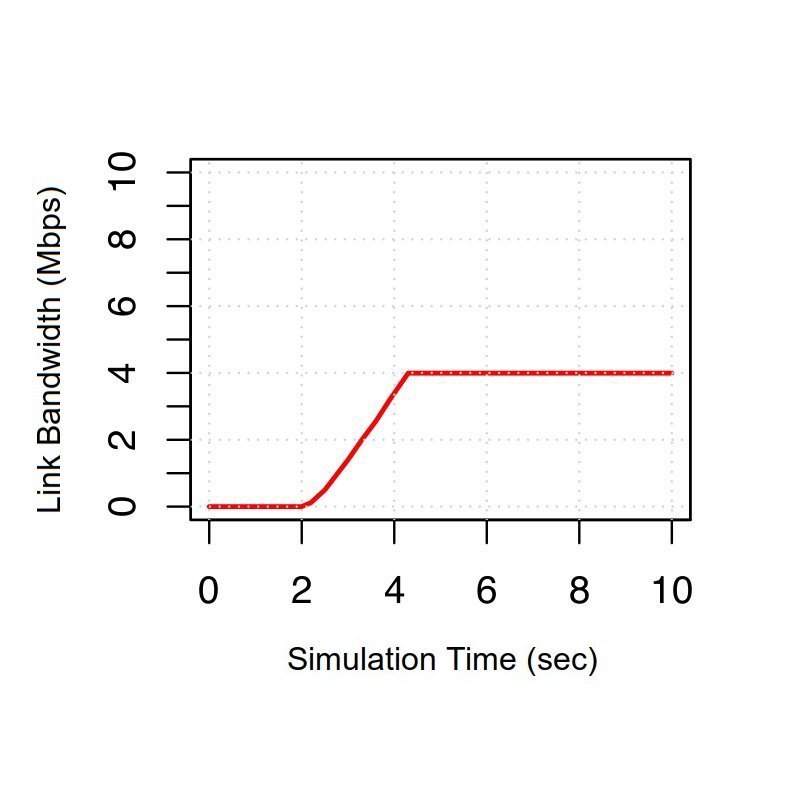}
		%\vspace{-0.8cm}
		\caption{Link between router 1 and router 2}
		\label{Figure:band2}
	\end{subfigure}
	\begin{subfigure}{.33\textwidth}
		\centering
		\includegraphics[scale=0.20]{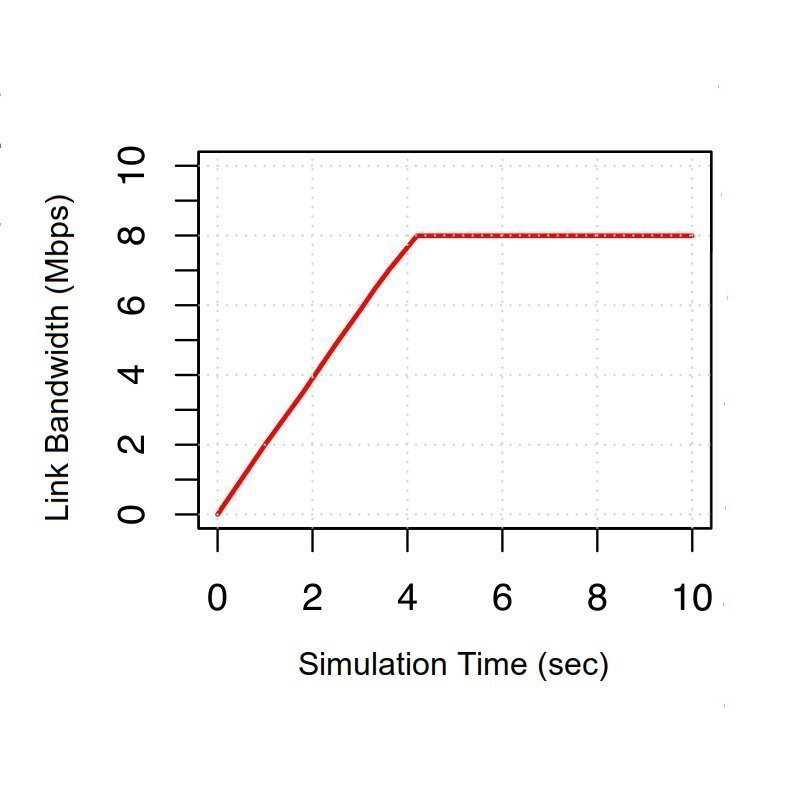}
		%\vspace{-0.8cm}
		\caption{Link between peer 1 and router 1}
		\label{Figure:band3}
	\end{subfigure}
	\caption{Utilization of Multiple Network Paths in Parallel by nTorrent for 1000 In-flight Interests (Scenario 1)}
	\label{Figure:topos}
\end{figure*}

\textbf {Scenario 1:} In Figure~\ref{Figure:topos}, we present the link utilization for the links between router 1 and router 3, router 1 and router 2, and peer 1 and router 1 of Figure~\ref{Figure:topology_routernodedeg3} respectively. Figure~\ref{Figure:band1} shows that the nTorrent forwarding strategy at router 1 first utilizes the link to router 3 and selects it as the primary link for Interest forwarding. This is due to the fact that the network paths toward peers 4 and 5 offer the lowest total data retrieval latency. After the primary link is fully utilized, our forwarding strategy at Router 1 is able to expand the forwarding of Interests from Peer 1 to the second available link, i.e., from router 1 to router 2 (Figure~\ref{Figure:band2}). 

Eventually, the strategy is able to keep the primary link fully utilized, while forwarding additional traffic toward router 2. In this way, peer 1 is able to achieve the best possible downloading speed based on our simulation setup, since the maximum bandwidth is the total of the bandwidth of the bottleneck links to peers 2, 3, 4, and 5 (Figure~\ref{Figure:band3}). Note that the nTorrent strategy at routers 2 and 3 first utilizes, in the same way as described above, the links with the shorter delay to the peers (e.g., link to peer 5 for router 3) and expands Interest forwarding to the second link available (e.g., link to peer 4 for router 3) after the primary link is fully utilized.

In Table~\ref{Table:scenario1}, we present the maximum reached download speed for a varying number of in-flight Interests, meaning transmitted Interests that are in the process of retrieving torrent data. The results show that as we increase the number of in-flight Interests, the nTorrent forwarding strategy is able to distribute them efficiently to multiple peers and increase the maximum download speed. Note that further increasing the number of in-flight Interests could cause congestion. In this case, our strategy sends back to the requester a NACK to indicate that the Interest sending rate needs to be reduced.

\begin{table}[]
\caption{Maximum Download Speed for a Varying Number of In-flight Interests}
\begin{tabular}{|c|c|}
\hline
\textbf{\begin{tabular}[c]{@{}c@{}}In-flight\\ Interests\end{tabular}} & \textbf{\begin{tabular}[c]{@{}c@{}}Download\\ Speed (Mbps)\end{tabular}} \\ \hline
100                                                                    & 0.84                                                                     \\ \hline
250                                                                    & 2.1                                                                      \\ \hline
500                                                                    & 4.2                                                                      \\ \hline
750                                                                    & 6.3                                                                      \\ \hline
1000                                                                   & $\sim$8                                                                  \\ \hline
\end{tabular}
\label{Table:scenario1}
%\vspace{-0.5cm}
\end{table}

\textbf {Scenario 2:} In Table~\ref{Table:scenario2}, we present the results for the second scenario of Section~\ref{sec:setup}, where peers 1, 2, and 3 (Figure~\ref{Figure:topology_routernodedeg4}) join at different points in time and download the torrent. Peer 1 starts downloading the torrent first and all the peers download the torrent data sequentially (starting from the first packet of the first file to the last packet of the last file, as mentioned in Section~\ref{sec:consumer}). As a result, peer 4 is the only one that can provide the data when requested by peer 1. That is why the results show that peer 1 downloads all the torrent data from peer 4, who is the original torrent seeder.

When peer 2 joins, data is available at peer 4 and peer 1. Given that the network latency to peer 1 is shorter, most of the torrent data is downloaded from peer 1. A small portion of the data is still downloaded from peer 4 due to the probing mechanism implemented by our forwarding strategy for the discovery of data potentially closer to the requester. When peer 3 joins, torrent data is available at all other peers. Given that the retrieval latency to peer 1 is the shortest, peer 3 downloads most of the data from peer 1. A small portion of the torrent is still downloaded from peer 2 and peer 4 due to the probing mechanism of the forwarding strategy.

\begin{table*}[]
\caption{Percent of Data Downloaded from Peers (Scenario 2)}
\begin{tabular}{|c|c|c|c|c|}
\hline
\textbf{Peer} & \textbf{\begin{tabular}[c]{@{}c@{}}Percent of torrent\\ data retrieved from\\ peer 1 (\%)\end{tabular}} & \textbf{\begin{tabular}[c]{@{}c@{}}Percent of torrent \\ data retrieved from \\ peer 2 (\%)\end{tabular}} & \textbf{\begin{tabular}[c]{@{}c@{}}Percent of torrent \\ data retrieved from \\ peer 3 (\%)\end{tabular}} & \textbf{\begin{tabular}[c]{@{}c@{}}Percent of torrent \\ data retrieved from \\ peer 4 (\%)\end{tabular}} \\ \hline
1             & 0                                                                                                       & 0                                                                                                         & 0                                                                                                         & 100                                                                                                       \\ \hline
2             & 95.6                                                                                                    & 0                                                                                                         & 0                                                                                                         & 4.4                                                                                                       \\ \hline
3             & 95.4                                                                                                    & 2.3                                                                                                       & 0                                                                                                         & 2.3                                                                                                       \\ \hline
\end{tabular}
\label{Table:scenario2}
%\vspace{-0.2cm}
\end{table*}

\textbf {Scenario 3:} For the scenario related to peer dynamics, our simulations showed that peer 1 initially retrieves torrent data from peer 2, since the network path between them offers the lowest latency. When peer 2 gets disconnected, the forwarding strategy at router 1 is able to identify that a number of consecutive Interests from peer 1 cannot get satisfied with torrent data through the path toward peer 2. As a result, router 1 starts forwarding the Interests of peer 1 toward peer 3 that now offers the lowest latency for data retrieval. When peer 3 gets disconnected, in the same way, router 1 is able to forward the Interests of peer 1 toward peer 4 that stays connected until the end of the torrent retrieval. 

\subsection {Further Framework Tests}

We have performed further tests to assess the correctness of our simulation framework. These tests include:

\emph{Flash crowd scenarios:} We tested our framework during a flash-crowd scenario, where multiple peers join over a short period of time and try to retrieve data initially served by a single peer (through a rocketfuel topology~\cite{spring2002measuring}). The results indicated that NDN in-network caching helps reduce the number of requests that needs to be satisfied by peers, since most of them are satisfied from the network in the case of multiple simultaneous data downloads.

\emph{Wireless file sharing scenarios:} We tested our framework with peers connected to Wi-Fi access points to simulate scenarios of laptops connected to home routers that share files over the Internet.

\emph{Scenarios with variable torrent sizes:} We have experimented with variable torrent sizes (ranging from a few MBs to a few GBs).

\emph{Variable Interest sending rates:} We have experimented with variable Interest sending rates to understand the trade-offs between congestion in the network and download time. 

\emph{Peer dynamics:} We have experimented with varying times of peers' arrivals and departures.

\section {Lessons Learned}
\label{sec:lessons}

Our ultimate goal is to merge our nTorrent simulation framework with the main branch of ns-3 and make it available to the ns-3 community for further experimentation. To achieve that, it is required that ndnSIM is first merged with ns-3, which is one of the plans of the ndnSIM team. In the rest of this section, we share the lessons we learned by working with a very large codebase that included the ndnSIM core, the NDN prototypes, and ns-3. These lessons also apply to users that develop new ns-3 modules or simply use ns-3 to run network simulations without adding new functionality, since ns-3 is also a large codebase consisting of a number of modules.

First of all, it is challenging to work with large codebases and there is a learning curve involved in the process. When the newly developed code does not have behave as expected, debugging is challenging, since bugs might not be associated with the developed code itself, but they might be bugs of the already existing large codebase. The ns-3 PyViz module (\url{https://www.nsnam.org/wiki/PyViz}) helped us better understand and debug the behavior of our framework and speculate what might go wrong. This module offers real-time visualization of the simulation execution.

Before starting our development, we cloned a specific version of ndnSIM and ns-3 and developed our own code based on this version. We later realized that we need to pull the latest changes of the ns-3 and ndnSIM repository frequently to minimize the number of changes introduced to ns-3 or/and ndnSIM, which could have unexpected side-effects on our code. For example, after pulling the changes of a new ndnSIM release, the API of the forwarding strategy had changed. Our code that was based on an older forwarding strategy API and thus needed to be updated.

We developed our framework based on the ndnSIM scenario template\footnote{\url{https://github.com/named-data-ndnSIM/scenario-template}}. We used stable versions of ndnSIM and ns-3 as system libraries, building and linking our code against them. This simplified the development of our framework, given that it did not require any changes in ndnSIM and ns-3. We only had to compile our framework code, without recompiling ndnSIM and ns-3.

We developed the nTorrent base library as a standalone software library, so that the provided abstractions can be directly re-used by new peer-to-peer applications in the future. To this end, we needed a way to build the library code as a part of our simulation framework. At this point, we realized the value of being able to link the library as a git submodule\footnote{\url{https://git-scm.com/book/en/v2/Git-Tools-Submodules}} to the rest of the nTorrent simulation framework. Apart from enabling the use of the library by other peer-to-peer applications as we mentioned above, the nTorrent simulation framework could also use a new library in the future by updating the submodule to point to this new library.

%\vspace{-0.2cm}

\section{Conclusions \& Future Work}
\label{sec:concl}

This paper describes the implementation of a simulation framework for peer-to-peer file sharing in NDN. This framework was implemented as a part of ndnSIM, the de facto NDN simulator, which is based on ns-3. We encourage the community to explore NDN research, experiment with the nTorrent simulation framework, and help us improve its design and implementation, as well as provide feedback for features that they would like us to develop. 

While the current implementation is a promising start, there are multiple areas of future work. First of all, we plan to make our implementation more modular through extensive code refactoring and support high-level scripting abstractions for the quick and easy instantiation of simulation scenarios. Another direction of future research work is to adapt the existing protocol to support wireless and mobile ad-hoc communication scenarios. Furthermore, we plan to merge our nTorrent simulation code with the main ns-3 codebase on GitHub (\url{https://github.com/nsnam/ns-3-dev-git}). Finally, we plan to investigate the scalability and trade-offs of the nTorrent design with more simulations, including more diverse topologies and various simulation setups, as well as explore hybrid NDN and Software-Defined Networking (SDN) solutions~\cite{mastorakis2018isa, argyropoulos2015control}.

%
% The acknowledgments section is defined using the "acks" environment (and NOT an unnumbered section). This ensures
% the proper identification of the section in the article metadata, and the consistent spelling of the heading.
%\vspace{-0.15cm}

%\begin{acks}

%We would like to thank the anonymous reviewers for their feedback and suggestions on improving the quality of this paper.

%\end{acks}

%\vspace{-0.35cm}

%
% The next two lines define the bibliography style to be used, and the bibliography file.
\bibliographystyle{ACM-Reference-Format}
\bibliography{sample-base}

% 
% If your work has an appendix, this is the place to put it.
%\appendix

\end{document}